\documentclass[a4paper]{article}

\usepackage{INTERSPEECH_v2}
\usepackage{float}
\usepackage{amsmath, graphicx, multirow, booktabs, cite}
  
\title{The Trajectory of Voice Onset Time with Vocal Aging}
\name{Chen Xuanda$^{1,2}$, Xiong Ziyu$^2$, Hu Jian$^3$}
\address{
  $^1$Department of Linguistics, Graduate School of Chinese Academy of Social Sciences, China\\
  $^2$Institute of Linguistics, Chinese Academy of Social Sciences, China\\
  $^3$School of Foreign Studies, Anhui University, China}
\email{chen.xuanda@foxmail.com, xiongziyu@163.com, hujianpost@126.com}
  
\begin{document}
  
  \maketitle
  \begin{abstract}

    Vocal aging, a universal process of human aging, can largely affect one's language use, possibly including some subtle acoustic features of one's utterances like Voice Onset Time. To figure out the time effects, Queen Elizabeth's Christmas speeches are documented and analyzed in the long-term trend. We build statistical models of time dependence in Voice Onset Time, controlling a wide range of other fixed factors, to present annual variations and the simulated trajectory. It is revealed that the variation range of Voice Onset Time has been narrowing over fifty years with a slight reduction in the mean value, which, possibly, is an effect of diminishing exertion, resulting from subdued muscle contraction, transcending other non-linguistic factors in forming Voice Onset Time patterns over a long time.
  
  \end{abstract}
  
  \noindent\textbf{Index Terms}: voice onset time, longitudinal variation, Christmas speech, phonetic corpora
  
\section{Introduction}
\label{sec:intro}

\begin{table*}[!t]
  \caption{Tokens in the dataset.}
  \label{table}
  \centering
  \begin{tabular}{ccccc}
    \toprule
    \textbf{Time Frame} & \textbf{Stop Type} & \textbf{Total Tokens} & \textbf{Deserted Tokens}  & \textbf{Retention Ratio}            \\
    \midrule
    \multirow{2}*{early} & voiced & 227 & \multirow{2}*{38} & \multirow{2}*{94.52\%} \\
    ~& -less & 470 &  ~&~  \\
    \multirow{2}*{mid} & voiced & 311 & \multirow{2}*{40} & \multirow{2}*{96.23\%} \\
    ~& -less & 763 &  ~&~  \\
    \multirow{2}*{late} & voiced & 463 & \multirow{2}*{89} & \multirow{2}*{94.17\%} \\
    ~& -less & 1064 &  ~&~  \\
    \bottomrule
  \end{tabular}
\end{table*}

Age-related voice changes have been validated in both spontaneous corpus and prompted speeches e.g. \cite{harrington2007evidence, hawkins2005formant}, while consonant, a more static segmental feature, is less influenced by the aging effects. Generally, Voice Onset Time, a classifier for stop consonants, is the duration of the period of time between the release of a plosive and the beginning of vocal fold vibration. Specifically, the voice onset time of voiceless stops is negatively correlated with speech rate, whereas the correlation is hard to detect for voiced stops. The place of articulation, as noticed, conditions the longest values for velar and the shortest for bilabial, following alveolar \cite{auzou2000voice, cho1999variation}. Additionally, in an experiment of reading narrative tasks, it is anticipated that prosodic strengthening entails longer voice onset time for stops with phrasal prominence or in phrase-initial position \cite{cole2007prosodic}. While these linguistic factors have been quite thoroughly studied, we are still unclear about some extra-linguistic factors that may be able to influence the value of Voice Onset Time like vocal aging or time dependence.

\subsection{Variation in voice onset time beyond linguistic factors}

Apart from the intrinsic properties of language, studies have revealed that the value of voice onset time can be socially conditioned, some of which even entail an interesting result that the length of voice onset time can be speaker-specific. In experiments with speaking rate control, individual differences in voice onset time still remain, which proves the idea that voice onset time differs as an intrinsic trait \cite{allen2003individual, alan2013phonetic}. In the light of sociolinguistics, gender, ethnicity and age were considered in the variation of voice onset time. In a study of African-American and Caucasian-American young and old, male and female speakers, compared with female and Caucasian-American speakers, males and African-Americans show more voicing of voiced stops while no significant variation between genders and ethnicities is observed among older subjects \cite{ryalls2004voice}. Docherty \cite{docherty2011variation} provides evidence that, in a fixed location, young speakers tend to produce longer voice onset time overall than the old, though the physiological basis has not been established. He notes another phenomenon as well that negative voice onset times for voiced stops are relatively infrequent among young speakers, which, he suspects, is consist with their tendency to adopt relatively long voice onset times for their voiceless stop realizations. Not every study supports the conclusion, however. Some researches claim that they can hardly find any significant age-related difference in voice onset time e.g. \cite{petrosino1993voice} or only complex interactions between ages, genders etc. e.g. \cite{torre2009age}.

Some of the results can result from failing to control variables like speech rate, a key factor that bring about hastened generalization \cite{morris2008voice}. Yet, speech rate cannot account for longer aspiration among the young speakers for, as studies have observed, elder speakers typically make slower utterances, which in turn should have caused longer aspiration and Voice Onset Time. Thus, two possible explanations are ahead of us: 1. vocal aging effects exclude variations in Voice Onset Time, that is, we cannot predict age only from Voice Onset Time itself and the experiments that observe longer Voice Onset Times among the young speakers and shorter among the old participants are conducted with some uncontrolled variable that largely affects the observation. 2. age and Voice Onset Time are linearly correlated and possibly the complex interactions among multiple miscellaneous variables have masked the simple linear dependence. The two hypotheses are left to verify in our experiment on large longitudinal corpus.

\subsection{Short-term and long-term shifting in voice onset time}

Variability in voice onset time does occur over time. Nielsen \cite{nielsen2011specificity} found in an experiment that speakers unconsciously lengthened voice onset time of word-initial stops when repeating a set of target words after exposure to productions with artificially extended voice onset times. It fits the prediction of the Speech Accommodation Theory, referring to speech alignment, or speech imitation, that naturally occurs in human contacts \cite{gallois2015communication}. Such changes, not only in experimental environment, are also found in real social contacts. Considering day-to-day variability in voice onset time of voiced and voiceless stops respectively, previous studies have built regression models on a corpus of spontaneous speech from eleven British subjects on a reality television show over a period of three months \cite{sonderegger2012phonetic}. The models verify daily fluctuations around the mean in voice onset time and a steady change in one’s mean voice onset time over a certain timescale is found in some of the subjects, which, unexpectedly, does not accumulate into long-term shift but bounces back to the previous stage. Yet, we do not reject the chance that the ubiquity of short-term change can be reconciled with the heterogeneity of long-term change on such basis of a three-month observation. To put it another way, short-term fluctuations persist on and hence, at a certain point, long-term trend shifts.

Very few studies indeed consider variability in voice onset time over long timescale while numerous accounts for historical shifts in stop aspiration really exist \cite{iverson1995aspiration}. Some measure the value of voice onset time in speech from the same group of individuals at a few time points, years apart while some do statistical analyses on large apparent-time corpora e.g. \cite{geiger2006reconstructing}. Geiger and Joseph found a slight reduction in voice onset time over time in German. The conclusion verifies that the length of voice onset time can change in both quality and quantity in the long run. Yet, these findings cannot provide strong evidences or clear depictions for the integral trajectory of voice onset time over one’s lifespan let alone proving whether it is universal as aging process or idiosyncratic associated merely with individual traits. At least, we believe an overall description for the trajectory of voice onset time with vocal aging can provide us with some insights into the local and global variation in stop consonants in general.

\section{Methodology}
\label{sec:typestyle}

Whether Voice Onset Time of stop consonants can be affected by vocal aging remains unknown though various acoustic representations have been studied from the longitudinal perspective. It is possibly due to the difficulty of controlling different variants such as speech rate, pitch range and so forth. In our experiment, we collect Christmas Speech as samples, and distinguish a list of fixed factors, which are non-time variables that greatly influence the length of Voice Onset Time. The statistical model is then able to capture time dependence only, so that the Voice Onset Time trajectory with vocal aging can be precisely depicted. On this basis and in an exploratory attempt, we address these research questions:
(1) Is the variation of Voice Onset Time of stop consonants vocal-aging sensitive or non-sensitive?
(2) What is the general tendency of the variation of Voice Onset Time and is a continuous describable trajectory possible?

\subsection{Data and corpus}
\label{ssec:subhead}

40 clips of the Queen's Christmas Speech (with total duration of 3 hours) have been collected ranging from 1953 to 2016, among which 23 years' clips, typically from 1954 to 1983, are either missing or beyond automatic identification in utterance. According to vocal maturity, three phases are split from the samples: [early stage, age 27 - 50, 9 clips]; [mid stage, age 50 - 70, 11 clips]; [late stage, age 70 - 90, 20 clips]. The forced alignment of transcription and signal is done by FAVE \cite{rosenfelder2011fave}, a python script to do phonetic transcription. After forced alignment five researchers with linguistic background are recruited to check the FAVE performance and make improvement. Automatic measurement of Voice Onset Time is done by AutoVOT \cite{sonderegger2012automatic} with following manual corrections.

Distinguished from original definition of pre-voicing in Voice Onset Time, only aspiration duration is considered in capturing the value, thus all the measurements exclude negative values. The phonologically voiced stops are regarded as ``burst + aspiration duration'' as previous studies \cite{sonderegger2012phonetic}. Some of the measurements are deserted if the spectrum is ambiguous while all the rest are labeled according to their linguistic and non-linguistic features, namely articulation position (labial, alveolar, velar), following phone type (vowel, consonant), syllable stress (stressed, unstressed), position in phrase (initial, non-initial) and speaking rate (syllables/sec). Finally, we obtained a list of log transformed measurements of Voice Onset Time in the dataset with corresponding feature labels.

\subsection{Model selection}
\label{ssec:subhead}

A linear mixed-effects model is described by the distribution of two vector-valued random variables: the response $y$, and the vector of random effects $\beta$. In a linear model, the distribution of the response is multivariate normal,
\begin{equation}
\centering
y \sim N(x\beta+\iota, \sigma^2W^{-1})
\end{equation}
where $N$ is the dimension of the response vector, $W$ is a diagonal matrix of known prior weights, $x$ is an n×p model matrix, $\beta$ is a p-dimensional coefficient vector, and $\iota$ is a vector of known prior offset terms. The parameters of the model are the coefficients $\beta$ and the scale parameter $\sigma$ \cite{bates2014fitting}. Given the current data, we have got 7 dimensions including stop id, Voice Onset Time value and above five fixed factors. Thus, we can fit the data into a linear mixed-effects model like,
\begin{equation}
\centering
fm <- lmer(log(vot) ~ F1 + ... + F5 + (1 | id), data)
\end{equation}
through which we fixed five static effects including all possible interactions and by-word random intercepts and slopes, and above all, we have captured residuals out of the fixed effects. Time dependence is then modeled in these residuals through generalized additive mixed models. The model provides a unified likelihood framework for parametric regression of a variety of over-dispersed and correlated outcomes just as residuals after fixing all static effects. GAMMs, for another advantage, allow for incorporation of two kinds of time trends: one is the random intercept of clips that may capture time trend; the other is the arbitrary smoothing function aross years, a non-linear smoother that captures time trend also. Such being the case, two time-dependent trajectories of voiced and voiceless stops can be extracted from the original data.

\section{Results}
\label{sec:majhead}

Annual variations in Voice Onset Time over 40 years have been captured along with two time-dependence models of voiced and voiceless stops separately, where the Voice Onset Time trend is linearly depicted. These are concerned with how to understand the symbols and trends in the rectangular coordinates and what we have truly perceived from the surface representations and underlying intricate relations.

\subsection{Annual variations of fluctuation in Voice Onset Time}
\label{ssec:subhead}

In order to see the fluctuation in Voice Onset Time with aging, we measured the time range of Voice Onset Time for each stop. Top 5\% and bottom 5\% values of voiced and voiceless stops have been extracted from the dataset and compared one by one within a certain year to entail a list of variations of gap differentials, each assigned with a new label of float “year.id” as the independent variable. A non-linear regression model, with accompanying smoothing, is then built in the variations to capture the tendency in general. The shadow that evenly distributed around the curve is the confidence interval and the inner curve indicates the tendency of annual gap differential variations.

\begin{figure}[htb]
  \begin{minipage}[b]{1.0\linewidth}
    \centering
    \centerline{\includegraphics[width=8.5cm]{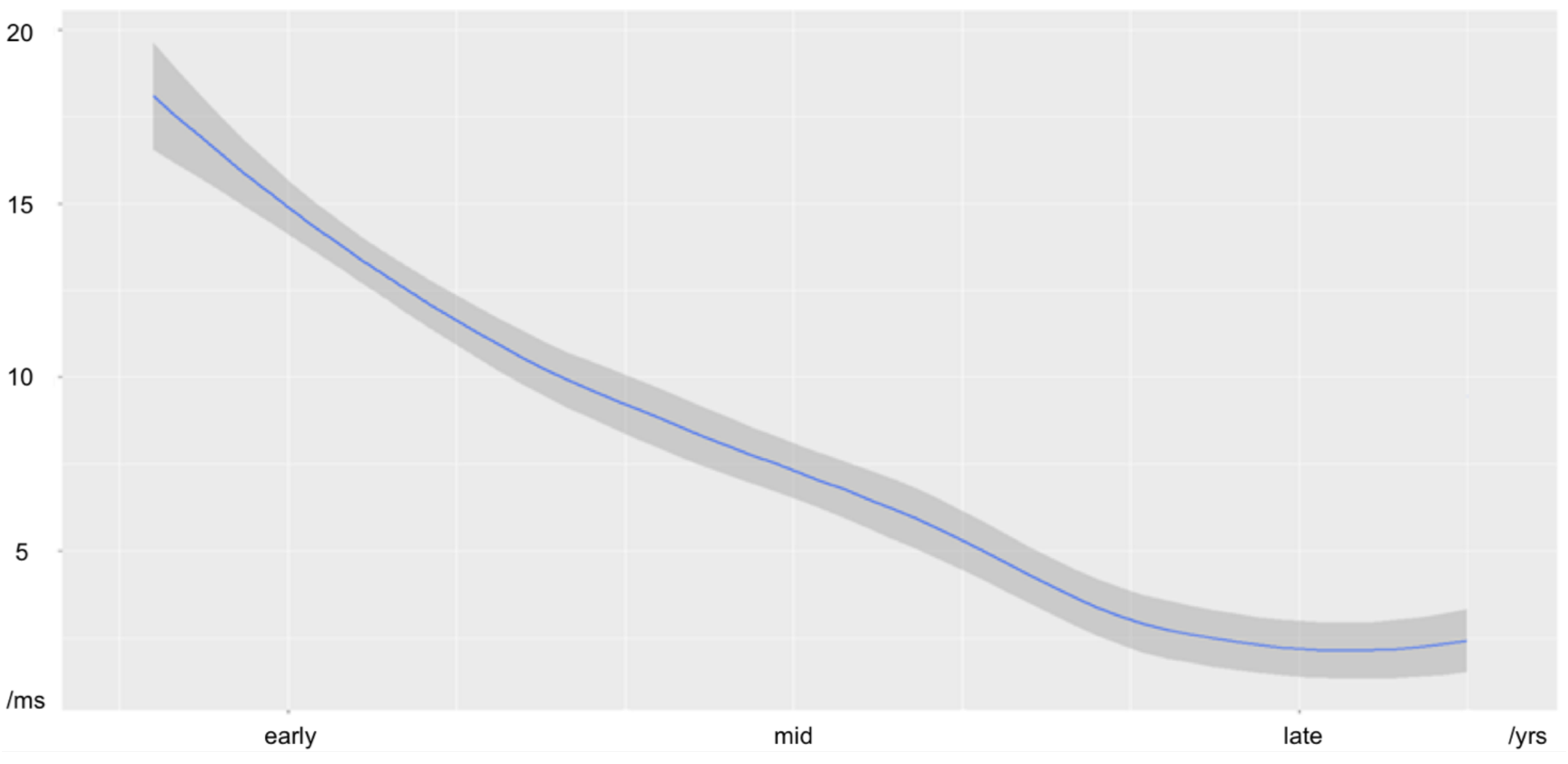}}
    \medskip
  \end{minipage}
\caption{Annual variations of fluctuation in voice onset time over 40 years.}
\label{fig:res}
\end{figure}

The tendency is close to linear until it comes to the time frame of late, which shifts from slight descending to nearly flat. The declining tendency is most obvious in the early period, and with statistical proof, the fastest declining occurs in the early period. In the mid period the slope gets smaller in the first part and in the next part, interestingly enough, the line is experiencing a sharp decrease with a similar slope ratio in the extreme point. Then quickly, the line gets into a stable state with a nearly zero slope ratio. Though the descending tendency is quite obvious in some periods, the whole fluctuation range is pretty limited from 0 ms to 20 ms. We do have witnessed larger gap like 30 ms, 40 ms or so, yet the quantity is so small that we regarded them as abnormal and have them ruled out from the modeling. In regard to the time span, the declining range might be unnoticeable.
 
\subsection{Trajectory trend simulation}
\label{ssec:subhead}

Time dependent simulation of Voice Onset Time is modeled in the residuals after deciding the fixed effects of certain linguistic factors. In figure 2, the left picture is the simulation result of voiced stops while the right is for the voiceless, with millisecond, the unit of measurement for the vertical axis. Likewise, precise years on horizontal axis are altered with three time frames as periods of early, mid and late, which are specially marked with the second quartile of the distribution. What calls for additional attention is that two figures share different coordinate distance on vertical axis so that it is inappropriate to have them compared without any distance transformation.

\begin{figure}[htb]
  \begin{minipage}[b]{1.0\linewidth}
    \centering
    \centerline{\includegraphics[width=8.5cm]{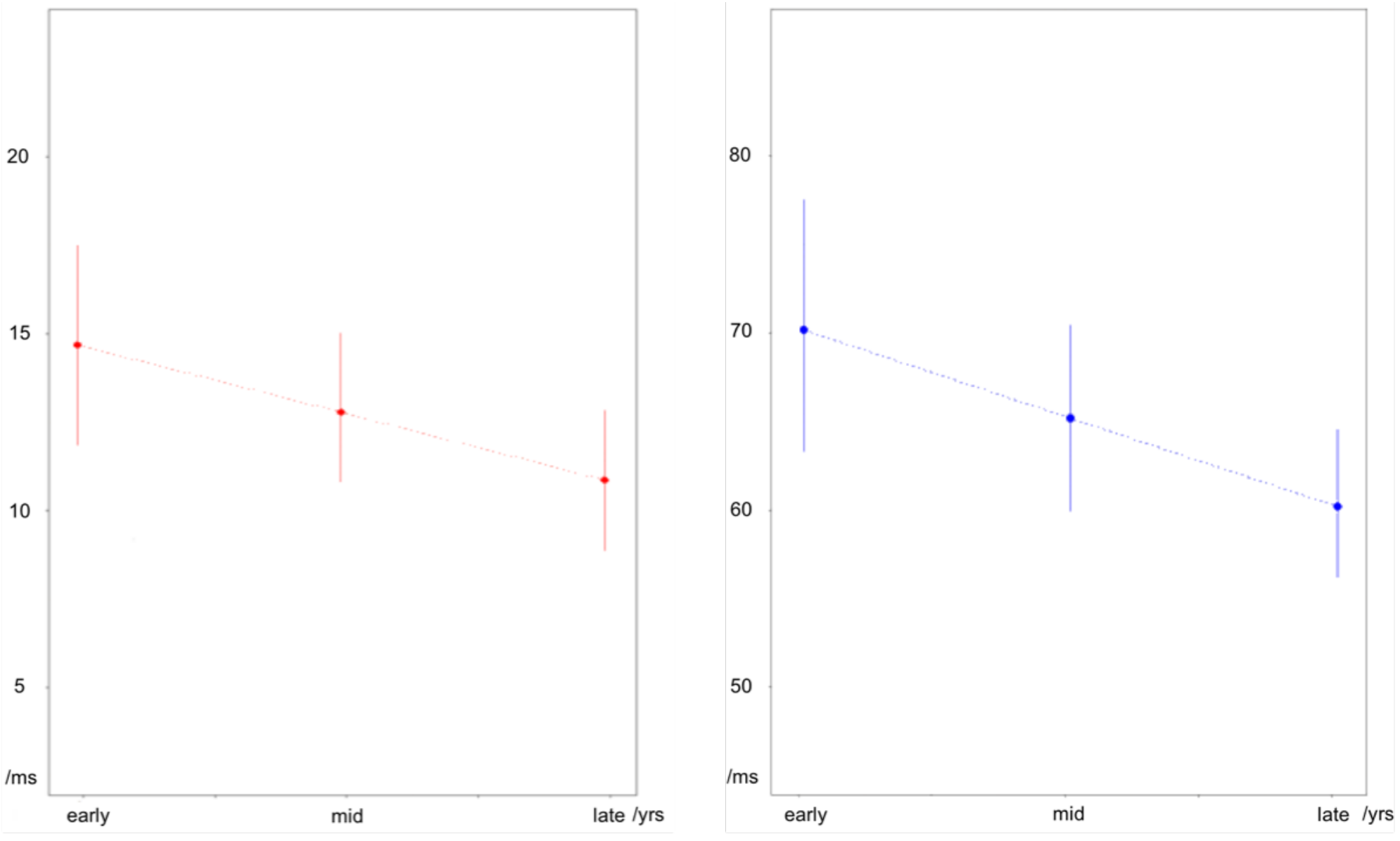}}
    \medskip
  \end{minipage}
\caption{Time-dependent simulation of voice onset time.}
\label{fig:res}
\end{figure}
       
It is revealed that Voice Onset Time of voiceless stop consonants is of greater declination than that of voiced stops. Coordinate distance concerned the difference can be even greater over the whole 40 years. Furthermore, the narrowing distribution of the second quartile is of varying declining speed as well. Specifically, voiceless stop consonants have been narrowing in Q2 distribution with a faster pace and till the late period, the span of Q2 is close to a half of that in the early period. Things are different for voiced stop consonants, however. The narrowing process is much slower and the narrowing degree is much smaller accordingly, with the largest span of 7 ms and the smallest of 4 ms. In general, the Voice Onset Time of both voiced and voiceless stops is similarly descending with time, or concretely, vocal aging.

\section{Discussion}
\label{sec:print}

To answer the questions we have addressed above, first of all, we can confirm that the Voice Onset Time does change with vocal aging, and even the variability range changes accordingly. Secondly, we observe from the time trend that the Voice Onset Time of both voiced stops and voiceless stops is descending yet of different slopes. Concerning physical aging process, it is assumed that longer aspiration is more sensitive to muscle contraction that produces speeches and such sensitivity is gradually numbed with age.

\subsection{Vocal aging and critical period}
\label{sssec:subsubhead}

Contrary to the conventional idea that consonants are generally more static over a long time, our computational modeling of sound change proves the dynamic nature of consonants, which, on a certain scale, varies regularly with time series and is easily predictable. Take Voice Onset Time in our dataset, over around 30 years of robust and strong vocal activity, the variability in Voice Onset Time drops from a high point with a comparatively high velocity while in the rest of the years, the velocity reduces to nearly zero, that is, for the last 10 years, the Voice Onset Time range for the subject maintains a stable condition. (What is interesting, yet, is a stable condition of range does not equal a stable value. As Figure 2 reveals, the Voice Onset Time of both voiced stops and voiceless stops roughly continue to decline, though with a low slope.) The figure cannot tell us an exact year when the range of Voice Onset Time begins to stop narrowing, but the curve indicates a period of time, the time frame of late, when the Voice Onset Time range meets a round inflexion. Can we regard the inflexion as a critical point of the variability? The answer might be no. Up to now, we have not found any strong physiological evidence of sharp vocal aging period as the period of vocal maturity for infants. Rather, some longitudinal studies offer strong support to the idea that vocal aging can be reversible, typically for damage from outside forces like smoking. Once the bad habits are stopped, vocal activity recovers to a well condition. 

While we are not convinced with the critical period from the evidence found in the experiment, the possibility of sound change hysteresis in late years is quite clear. In the late period, the gap between voiced and voiceless stops is narrowed and the trend becomes flat. Reasonably speaking, we assume such phenomenon is associated with the condition of vocal tract. The flat part of the variability curve also indicates equilibrium between vocal aging and reduction of Voice Onset Time, along with which the muscle contraction may stop weakening or just keep diminishing at an unnoticeable pace. It is still unknown, however, whether such phenomenon is speaker-specific or universally applicable, for the vocal aging process may vary from person to person.

\subsection{The descending trend of Voice Onset Time}
\label{sssec:subsubhead}

Five types of trajectory were presumed before the conduction of experiments, namely a) linear increase; b) linear decrease; c) fluctuating increase; d) fluctuating decrease; e) stable horizontal line. All five trajectories are considered time-dependent only, regardless of by-word random dependence. The result, as indicated in Figure 2, supports prediction b, though voiced and voiceless stops share different slopes of descending trend. The descending trend is consistent with the previous analyses and predictions as we have discussed earlier. Yet, we tend to deny that speaking rate, having influence on the value of Voice Onset Time, declines with age. Rather, as explained above, more chances are that the declining of muscle contraction prevents stronger aspiration for stops. This is where we can find more evidence from linguistic and physiological studies. In the first place, reviewing the production of Voice Onset Time, it is defined as the length of time that passes between the release of a stop consonant and the onset of voicing, the vibration of the vocal folds, or, for some, periodicity. The release of a stop matters in measuring the Voice Onset Time, for its duration directly influences the final value of the measurement. It is especially true of our experiment, which ignores the voicing that begins during the period of articulatory closure. For another, the release, if we are looking for physiological support, is closed tied to the initial force and the uphold of muscular tension (the situation now we have is the duration of Voice Onset Time may sometimes be easier to measure than aerodynamic, articulatory parameters). 

\begin{figure}[htb]
  \begin{minipage}[b]{1.0\linewidth}
    \centering
    \centerline{\includegraphics[width=8.5cm]{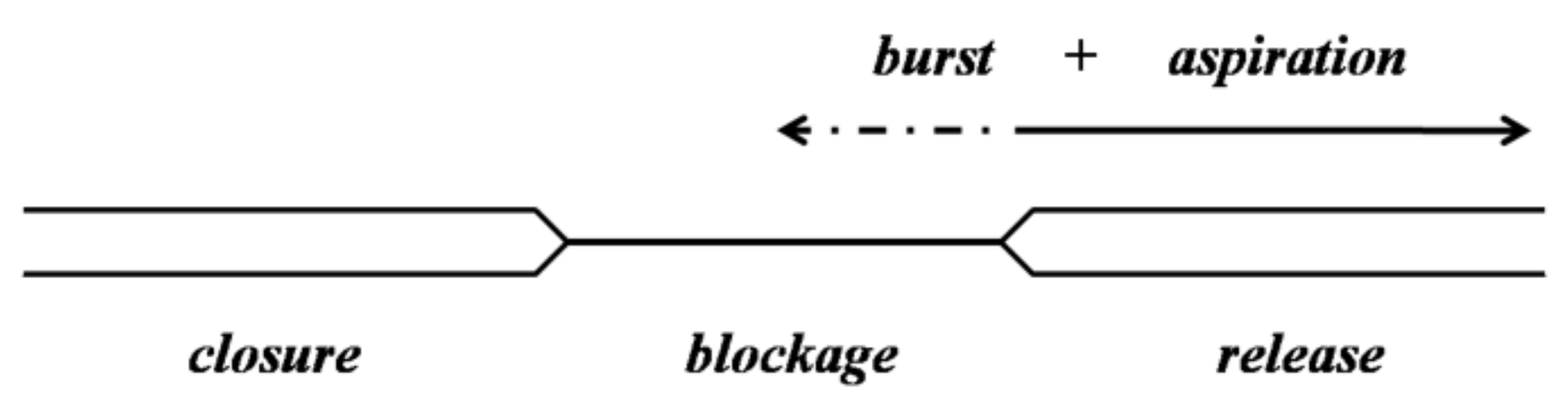}}
    \medskip
  \end{minipage}
\caption{Graphical representation of voice onset time.}
\label{fig:res}
\end{figure}

A graphical representation of Voice Onset Time mechanism and measurement reveals a linear positive relation between release gesture and airflow production, or in other words, aspiration. Previous study found subjects gradually lose control on oral movement and voice stability, and more often, the loss of some vocal functions deteriorates acoustic voice signals and adds vocal roughness as well \cite{verdonck2004vocal}. Such being the case, we could account for the reduced value of Voice Onset Time with physiological evidence: with loss of control on vocal muscles, subjects are more likely produce targetless sounds, ones that deviates from the expected target and fails to satisfy full pronunciations, including the habitual aspiration of stop consonants. In the meantime, vocal fragility cannot uphold increasing tension to produce longer aspiration as simulated in time dependent model. Now another question is, however, whether the reduced Voice Onset Time is continuously declining. The descending trend seems to linearly continue for both voiced and voiceless stops while the variation gap has been narrowing and even reached to equilibrium over the past 40 years. It is hard to decide whether or when the declining would cease just based on the given data and varying reference systems. Yet a possible inference from the current experiment lies in the cause of time dependent Voice Onset Time reduction: since we have decided the underlying biological factor, it is reasonable to convert the question of how is the following descending trend to that of how is the vocal aging process.

\section{Conclusion}
\label{sec:page}

The experiment above successfully describes the trajectory of Voice Onset Time with vocal aging and proves the existence of subtle variability over time. It is assumed that the narrowing annual variation and the slight reduction in the mean Voice Onset Time can be traced back to subdued muscle contraction with vocal aging, and the effect transcends other non-linguistic factors in forming Voice Onset Time patterns over a long time. Yet, still, we are not able to offer a reasonable explanation for the transcendence of subdued muscle contraction over other non-linguistic factors. Looking back on the whole experiment, we have found some subtle errors like mistreat to the void values of missing clips in sample modeling and a limited set of fixed factors that might influence the result of time dependence modeling. Thus, in the future, we are considering altering the modeling methods and expanding the speech data in an attempt to capture the refined relationship between the variability of Voice Onset Time and vocal aging.
  
  \bibliographystyle{IEEEtran}
  
  \bibliography{mybib}

% Generated by IEEEtran.bst, version: 1.14 (2015/08/26)
\begin{thebibliography}{10}
\providecommand{\url}[1]{#1}
\csname url@samestyle\endcsname
\providecommand{\newblock}{\relax}
\providecommand{\bibinfo}[2]{#2}
\providecommand{\BIBentrySTDinterwordspacing}{\spaceskip=0pt\relax}
\providecommand{\BIBentryALTinterwordstretchfactor}{4}
\providecommand{\BIBentryALTinterwordspacing}{\spaceskip=\fontdimen2\font plus
\BIBentryALTinterwordstretchfactor\fontdimen3\font minus
  \fontdimen4\font\relax}
\providecommand{\BIBforeignlanguage}[2]{{%
\expandafter\ifx\csname l@#1\endcsname\relax
\typeout{** WARNING: IEEEtran.bst: No hyphenation pattern has been}%
\typeout{** loaded for the language `#1'. Using the pattern for}%
\typeout{** the default language instead.}%
\else
\language=\csname l@#1\endcsname
\fi
#2}}
\providecommand{\BIBdecl}{\relax}
\BIBdecl

\bibitem{harrington2007evidence}
J.~Harrington, ``Evidence for a relationship between synchronic variability and
  diachronic change in the queen’s annual christmas broadcasts,''
  \emph{Laboratory phonology}, vol.~9, pp. 125--143, 2007.

\bibitem{hawkins2005formant}
S.~Hawkins and J.~Midgley, ``Formant frequencies of rp monophthongs in four age
  groups of speakers,'' \emph{Journal of the International Phonetic
  Association}, vol.~35, no.~2, pp. 183--199, 2005.

\bibitem{auzou2000voice}
P.~Auzou, C.~Ozsancak, R.~J. Morris, M.~Jan, F.~Eustache, and D.~Hannequin,
  ``Voice onset time in aphasia, apraxia of speech and dysarthria: a review,''
  \emph{Clinical Linguistics \& Phonetics}, vol.~14, no.~2, pp. 131--150, 2000.

\bibitem{cho1999variation}
T.~Cho and P.~Ladefoged, ``Variation and universals in vot: evidence from 18
  languages,'' \emph{Journal of phonetics}, vol.~27, no.~2, pp. 207--229, 1999.

\bibitem{cole2007prosodic}
J.~Cole, H.~Kim, H.~Choi, and M.~Hasegawa-Johnson, ``Prosodic effects on
  acoustic cues to stop voicing and place of articulation: Evidence from radio
  news speech,'' \emph{Journal of Phonetics}, vol.~35, no.~2, pp. 180--209,
  2007.

\bibitem{allen2003individual}
J.~S. Allen, J.~L. Miller, and D.~DeSteno, ``Individual talker differences in
  voice-onset-time,'' \emph{The Journal of the Acoustical Society of America},
  vol. 113, no.~1, pp. 544--552, 2003.

\bibitem{alan2013phonetic}
C.~Alan, C.~Abrego-Collier, and M.~Sonderegger, ``Phonetic imitation from an
  individual-difference perspective: Subjective attitude, personality and
  “autistic” traits,'' \emph{PloS one}, vol.~8, no.~9, p. e74746, 2013.

\bibitem{ryalls2004voice}
J.~Ryalls, M.~Simon, and J.~Thomason, ``Voice onset time production in older
  caucasian-and african-americans,'' \emph{Journal of Multilingual
  Communication Disorders}, vol.~2, no.~1, pp. 61--67, 2004.

\bibitem{docherty2011variation}
G.~J. Docherty, D.~Watt, C.~Llamas, D.~Hall, and J.~Nycz, ``Variation in voice
  onset time along the scottish-english border,'' in \emph{Proceedings of the
  17th International Congress of Phonetic Sciences}, 2011, pp. 591--594.

\bibitem{petrosino1993voice}
L.~Petrosino, R.~D. Colcord, K.~B. Kurcz, and R.~J. Yonker, ``Voice onset time
  of velar stop productions in aged speakers,'' \emph{Perceptual and motor
  skills}, vol.~76, no.~1, pp. 83--88, 1993.

\bibitem{torre2009age}
P.~Torre and J.~A. Barlow, ``Age-related changes in acoustic characteristics of
  adult speech,'' \emph{Journal of communication disorders}, vol.~42, no.~5,
  pp. 324--333, 2009.

\bibitem{morris2008voice}
R.~J. Morris, C.~R. McCrea, and K.~D. Herring, ``Voice onset time differences
  between adult males and females: Isolated syllables,'' \emph{Journal of
  Phonetics}, vol.~36, no.~2, pp. 308--317, 2008.

\bibitem{nielsen2011specificity}
K.~Nielsen, ``Specificity and abstractness of vot imitation,'' \emph{Journal of
  Phonetics}, vol.~39, no.~2, pp. 132--142, 2011.

\bibitem{gallois2015communication}
C.~Gallois and H.~Giles, ``Communication accommodation theory,'' \emph{The
  international encyclopedia of language and social interaction}, 2015.

\bibitem{sonderegger2012phonetic}
M.~Sonderegger, \emph{Phonetic and phonological dynamics on reality
  television}.\hskip 1em plus 0.5em minus 0.4em\relax University of Chicago,
  2012.

\bibitem{iverson1995aspiration}
G.~K. Iverson and J.~C. Salmons, ``Aspiration and laryngeal representation in
  germanic,'' \emph{Phonology}, vol.~12, no.~3, pp. 369--396, 1995.

\bibitem{geiger2006reconstructing}
S.~R. Geiger and J.~C. Salmons, ``Reconstructing variation at shallow time
  depths,'' \emph{Variation and reconstruction}, vol. 268, p.~37, 2006.

\bibitem{rosenfelder2011fave}
I.~Rosenfelder, J.~Fruehwald, K.~Evanini, and J.~Yuan, ``Fave (forced alignment
  and vowel extraction) program suite,'' \emph{URL http://fave. ling. upenn.
  edu}, 2011.

\bibitem{sonderegger2012automatic}
M.~Sonderegger and J.~Keshet, ``Automatic measurement of voice onset time using
  discriminative structured prediction a,'' \emph{The Journal of the Acoustical
  Society of America}, vol. 132, no.~6, pp. 3965--3979, 2012.

\bibitem{bates2014fitting}
D.~Bates, M.~M{\"a}chler, B.~Bolker, and S.~Walker, ``Fitting linear
  mixed-effects models using lme4,'' \emph{arXiv preprint arXiv:1406.5823},
  2014.

\bibitem{verdonck2004vocal}
I.~M. Verdonck-de Leeuw and H.~F. Mahieu, ``Vocal aging and the impact on daily
  life: a longitudinal study,'' \emph{Journal of Voice}, vol.~18, no.~2, pp.
  193--202, 2004.

\end{thebibliography}
  
  \end{document}